\documentclass[reprint,
superscriptaddress,
amsmath,
amssymb,
aps,
pra,
floatfix
]{revtex4-1} % revtex4-2 does not work using Overleaf

\usepackage{graphicx}% Include figure files
\usepackage{physics}
\usepackage{dcolumn}% Align table columns on decimal point

\usepackage{hyperref} % Add hypertext capabilities
\hypersetup{%
     colorlinks   = true,
     breaklinks   = true,
     linkcolor    = {blue},
     anchorcolor  = {blue},
     filecolor    = {blue},
     citecolor    = {blue},
     urlcolor     = {blue},
}
\usepackage{float}

\begin{document}

\title{Applying the Quantum Approximate Optimization Algorithm to \\ the Tail Assignment Problem}

\author{Pontus Vikst{\aa}l}
\email[e-mail: ]{vikstal@chalmers.se}
\affiliation{Wallenberg Centre for Quantum Technology, Department of Microtechnology and Nanoscience, Chalmers University of Technology, 412 96 Gothenburg, Sweden}

\author{Mattias Gr{\"o}nkvist}
\affiliation{Jeppesen Systems AB, 411 03 Gothenburg, Sweden.}

\author{Marika Svensson}
\affiliation{Wallenberg Centre for Quantum Technology, Department of Microtechnology and Nanoscience, Chalmers University of Technology, 412 96 Gothenburg, Sweden}
\affiliation{Jeppesen Systems AB, 411 03 Gothenburg, Sweden.}

\author{Martin Andersson}
\affiliation{Jeppesen Systems AB, 411 03 Gothenburg, Sweden.}

\author{G{\"o}ran Johansson}
\affiliation{Wallenberg Centre for Quantum Technology, Department of Microtechnology and Nanoscience, Chalmers University of Technology, 412 96 Gothenburg, Sweden}

\author{Giulia Ferrini}
\affiliation{Wallenberg Centre for Quantum Technology, Department of Microtechnology and Nanoscience, Chalmers University of Technology, 412 96 Gothenburg, Sweden}

\date{\today}

\begin{abstract}
Airlines today are faced with a number of large scale scheduling problems. One such problem is the Tail Assignment problem, which is the task of assigning individual aircraft to a given set of flights, minimizing the overall cost. Each aircraft is identified by the registration number on its tail fin. In this article, we simulate the Quantum Approximate Optimization Algorithm (QAOA) applied to instances of this problem derived from real world data. The QAOA is a variational hybrid quantum-classical algorithm recently introduced and likely to run on near-term quantum devices. The instances are reduced to fit on quantum devices with 8, 15 and 25 qubits. The reduction procedure leaves only one feasible solution per instance, which allows us to map the Tail Assignment problem onto the Exact Cover problem. We find that repeated runs of the QAOA identify the feasible solution with close to unit probability for all instances. Furthermore, we observe patterns in the variational parameters such that an interpolation strategy can be employed which significantly simplifies the classical optimization part of the QAOA. Finally, we empirically find a relation between the connectivity of the problem graph and the single-shot success probability of the algorithm.
\end{abstract}
\maketitle
%=============================================
%                INTRODUCTION
%=============================================
\section{\label{sec:Introduction}Introduction}
Real world planning and scheduling problems typically require heuristic algorithms, which is also the case for the Tail Assignment problem. The problem is to assign a set of flights to a set of aircraft in order to create a feasible flight schedule for an airline, while minimizing the overall cost~\cite{Gronkvist}.

Recently, quantum computing hardware has reached the regime where it is possible to run quantum algorithms which are hard to simulate on classical hardware, even considering the world's largest supercomputer~\cite{Arute2019}. This motivates the search for a heuristic quantum algorithm for solving the Tail Assignment problem. A promising approach for this is the Quantum Approximate Optimization Algorithm (QAOA)~\cite{Farhi2014}, which is a heuristic hybrid quantum-classical algorithm designed for solving combinatorial optimization problems. Since the algorithm was first proposed by Farhi et al.~\cite{Farhi2014} it has been an active area of research interest~\cite{Crooks2018,willsch2019,shaydulin2019,alam2019,xue2019,Brandao2018,Farhi2019}, mainly because of its promising possibility to run on a near term Noisy Intermediate Scale Quantum (NISQ) device. An important open question is whether a quantum computer in general can provide advantages with regards to such classically hard combinatorial optimization problems. Recent studies have indicated that QAOA can have a quadratic Grover type speed up for state transfer and unstructured search problems \cite{niu2019,Jiang2017}. Although these results are promising, the performance is largely unknown for QAOA with respect to real world optimization problems. 

Here we present, to our knowledge, the first results for QAOA when applied to a real world aircraft assignment problem. We perform numerical simulations of an ideal quantum computer to investigate the performance of QAOA for solving the simplified case of the Tail Assignment problem where all costs are equal to \emph{zero}. This simplified case can be mapped onto the Exact Cover problem ~\cite{parmentier2019aircraft}. In this context, we note that the solution of random instances of the Exact Cover and of its restricted version Exact Cover by 3-sets on a quantum annealer has been considered in Refs.~\cite{Farhi2001,Young2010,Altshuler2010,Choi2010,Wang2016,Grass2019}. QAOA for Exact Cover has recently been executed on a $2$-qubit quantum computer in a proof-of-principle experiment by some of the authors of the present paper, and collaborators~\cite{Andreas2019}.

The paper is organized as follows. In Sec.~\ref{sec:TAS}, we introduce the Tail Assignment problem, and we explain how we extract the Exact Cover instances that we analyze in this work. In Sec.~\ref{sec:QAOA}, we review the QAOA and explain how it can be utilized to solve the Exact Cover problem. Then, in Sec.~\ref{sec:Results} we present numerical results of the performance of QAOA with respect to the Tail Assignment problem-extracted instances of Exact Cover for three different problem sizes. Specifically, we look at the dependence of the success probability as a function of the algorithm iteration level $p$ and of the problem size. Finally, in Sec.~\ref{sec:Conclusions} we present what implications these results might have for solving the Tail Assignment problem.
%=============================================
%           TAIL ASSIGNMENT PROBLEM
%=============================================
\section{\label{sec:TAS}The Tail Assignment Problem}
Airlines are daily confronted with several complicated large-scale planning problems involving many different resource types such as passengers, crew, aircraft, maintenance and ground staff. The typical airline planning process~\cite{Jacobs2017Chapter2A} is a sequential process which starts with the construction of a timetable, followed by a number of aircraft and crew planning steps. These steps are all large scale optimization problems and have different objectives, but the overall goal is to maximize profit, safety and crew satisfaction while minimizing the potential for disruptions. At the same time a large number of complex regulatory, operational and quality constraints must be satisfied. 

The Tail Assignment problem~\cite{Gronkvist} is one of the fleet planning problems where the goal is to decide which individual aircraft (or tail, from the aircraft tail identification number) should operate each flight. A set of flights operated in sequence by the same aircraft is called a \textit{route}. In order for a route to be considered legal to operate, it needs to satisfy a number of constraints. For example, the buffer time between the arrival of a flight and the departure of the next flight in the route (the \textit{turn time}) must be above a certain threshold, called the minimum turn time. The minimum turn time can depend on the type of flights involved (domestic/international), the airport, the time of day and possibly even the individual aircraft characteristics. Another type of constraint is a destination restriction, which prohibits specific aircraft from visiting certain airports, for example due to limited engine thrust combined with short runways. Curfew restrictions are timed restrictions, typically limiting noisy aircraft from operating during night hours at centrally placed airports. Finally, routes must satisfy a number of long and short term maintenance constraints. This typically means that the aircraft must regularly visit some airport with a maintenance facility for long enough to perform maintenance.

Now, let $F$ denote the set of flights, $T$ the set of tails and $R$ the set of all legal routes. Denote by $c_r$ the cost of route $r \in R$ and by $C_f$ the cost of leaving flight $f$ unassigned. The route cost can for example indicate how robust the route is with respect to disruptions, what the fuel cost is for the route, or a combination of several different criteria. Let $a_{fr}$ be $1$ if flight $f$ is covered by route $r$ and $0$ otherwise, and let $b_{tr}$ be $1$ if route $r$ uses tail $t$ and $0$ otherwise. The decision variable $x_r$ is $1$ if route $r$ should be used in the solution, and $0$ otherwise. The variables $u_f$ and $v_t$ are $1$ if flight $f$ is left unassigned or tail $t$ is unused, respectively, and $0$ otherwise. The Tail Assignment problem can now be formulated as
\begin{alignat}{2}
   & \text{minimize\ } & \sum_{r \in R}c_r x_r + \sum_{f \in F}C_f u_f, & \label{eq:MinCost} \\
    & \text{subject to\ } & \sum_{r \in R} a_{fr}x_r + u_f &= 1, \quad \forall f \in F, \label{eq:FlightCover} \\
    & & \sum_{r \in R} b_{tr}x_r + v_t &= 1, \quad \forall t \in T, \label{eq:TailCover} \\
    & & x_r, u_f, v_t &\in \{0, 1\} \label{eq:Integrality}
\end{alignat}
The objective \eqref{eq:MinCost} is to minimize the total cost of the selected routes, subject to constraints \eqref{eq:FlightCover} ensuring that each flight is assigned to exactly one route and constraints \eqref{eq:TailCover} ensuring that each tail is used at most once. Flights can be left unassigned at a cost $C_f$, but that cost is typically very high compared to the route costs. Not using an aircraft does not come with any penalty cost. The model is an example of a Set Partitioning problem, which is NP-hard~\cite{GareyJohnson79}.
%=============================================
%                 SOLVING TAS
%=============================================
\subsection{\label{sec:SolvingTAS}Solving the Tail Assignment Problem}
Clearly, the number of legal routes for a Tail Assignment instance increases exponentially with the number of flights. Since the model presented above requires all the legal routes to be enumerated, it only works for small instances. The solution method traditionally used for these types of models is \textit{column generation}~\cite{Gronkvist}. Column generation starts from some initial solution and uses information from the linear programming dual problem to dynamically generate new variables (columns in the constraint matrix) which are known to potentially improve the current solution. In the Tail Assignment case, the problem of generating improving variables turns out to be a resource constrained shortest path problem. Given mild conditions on the variable generation step, the column generation process can be shown to guarantee optimality for the LP relaxation of the problem, i.e. without the integrality conditions Eq.~\eqref{eq:Integrality}. To find an optimal solution for the full problem including the integrality conditions, column generation must be combined with tree search. The combination of tree search and column generation is often called \textit{branch-and-price} \cite{Barnhart96branch-and-price:column}.
%=============================================
%               INSTANCE EXTRACTION
%=============================================
\subsection{\label{sec:TASinstance-generation}Instances extraction}
For the purposes of this article, given the current capability of quantum computers, we will focus on Tail Assignment instances where we have artificially limited the number of routes. The instances have originally been solved using a branch-and-price heuristic, and we have randomly selected a number of routes from the set of all generated routes to create instances of specific sizes. The solution found by the branch-and-price heuristic is always included, so we know that all instances have a solution with all flights assigned. This means that we can skip the $u_f$ variables in the model. We also have uniquely assigned start flights for each aircraft, which means that constraints Eq.~\eqref{eq:TailCover} can be omitted. Finally, in the remainder of this article we will focus on the decision version of the Tail Assignment problem where the goal is to find any solution satisfying all the constraints, disregarding the costs $c_r$. This decision version of the Set Partitioning problem is called the Exact Cover problem, it is know to be NP-complete~\cite{Karp1972}, and can be expressed as the following optimization problem:
\begin{alignat}{2}
   & \text{minimize\ } & 0 & \label{eq:MinCostExactCover} \\
    & \text{subject to\ } & \sum_{r \in R} a_{fr}x_r &= 1, \quad \forall f \in F, \label{eq:FlightCoverExactCover} \\
    & & x_{r} &\in \{0, 1\}, \label{eq:IntegralityExactCover}
\end{alignat}
where the minimization on $0$ is left to recall that this formulation stems from the Tail Assignment problem, where we neglect the costs in Eq.(\ref{eq:MinCost}).
Despite the simplification introduced, the Exact Cover problem is still very relevant for the study of Tail Assignment as  many airlines, including for example Air France, consider the Tail Assignment problem to be a pure feasibility problem \cite{Parmentier_2020}. 
%=============================================
%               QAOA APPLIED TO TAS
%=============================================
\section{\label{sec:QAOA}QAOA applied to the Tail Assignment problem}
A large class of NP-complete optimization problems including the Exact Cover (and even many NP-hard problems) can naturally be expressed as the problem of finding the ground state, or minimum energy configuration, of a quantum Ising Hamiltonian~\cite{Lucas2014}
\begin{equation}
    \label{eq:CostFunction}
    \hat{H}_C = \sum_{i<j}J_{ij}\hat{\sigma}^z_i\hat{\sigma}^z_j+\sum_{i=1}^nh_i\hat{\sigma}^z_i.
\end{equation}
We will refer to this quantum Ising Hamiltonian as a cost Hamiltonian. In this section, we derive explicitly the cost Hamiltonian corresponding to the Exact Cover problem expressed by Eq.~\eqref{eq:FlightCoverExactCover} and \eqref{eq:IntegralityExactCover}. Later, we recall the QAOA algorithm, and in particular how it makes use of the cost Hamiltonian for finding its minimum energy configuration.
%=============================================
%              ISING FORMULATION
%=============================================
\subsection{\label{app:IsingFormulation}Ising formulation of the Exact Cover problem}
Consider the formulation of the Exact Cover problem presented in Eq.~\eqref{eq:FlightCoverExactCover} and \eqref{eq:IntegralityExactCover}. By subtracting $1$ from both sides of Eq.~\eqref{eq:FlightCoverExactCover} and squaring the expression an energy function formulation is obtained:
\begin{equation}
    \label{eq:PenalizedFlightCover}
    \mathcal{E}(s_1,\ldots,s_{|R|})=\sum_{f=1}^{|F|}\left(\sum_{r=1}^{|R|}a_{fr}x_r-1\right)^2.
\end{equation}
Here $|R|$ and $|F|$ denote the cardinality of $R$ and $F$, respectively. We see that all constraints are satisfied if the energy \eqref{eq:PenalizedFlightCover} is equal to \emph{zero}.

By replacing the binary variables $x_r\in\{0,1\}$ with spin variables $s_r \in\{-1,1\}$ as
\begin{equation}
    x_r=\frac{s_r+1}{2},
\end{equation}
and expanding the square of Eq.~\eqref{eq:PenalizedFlightCover} we obtain the \emph{Ising energy function} for the Exact Cover problem
\begin{align}
  \mathcal{E}(s_1,\ldots,s_{|R|}) =  &  \sum_{f=1}^{|F|}\left(\sum_{r=1}^{|R|}a_{fr}\frac{s_r+1}{2}-1\right)^2=
    \nonumber \\
 + &   \frac{1}{4}\sum_{f=1}^{|F|}\sum_{r=1}^{|R|}\sum_{r'=1}^{|R|}a_{fr}a_{fr'}s_rs_{r'} \nonumber \\
 + &   \frac{1}{2}\sum_{f=1}^{|F|}\sum_{r=1}^{|R|}a_{fr}s_r\left(\sum_{r'=1}^{|R|}a_{fr'}-2\right) \nonumber \\
 + &   \frac{1}{4}\sum_{f=1}^{|F|}\left(\sum_{r=1}^{|R|}a_{fr}-2\right)^2.
\end{align}
By defining $J_{rr'}$ as
\begin{equation}
    \label{eq:Interaction}
    J_{rr'}\equiv\frac{1}{2}\sum_{f=1}^{|F|}a_{fr}a_{fr'},
\end{equation}
and $h_r$ as
\begin{equation}
    \label{eq:MagneticField}
    h_r\equiv\frac{1}{2}\sum_{f=1}^{|F|}a_{fr}\left(\sum_{r'=1}^{|R|}a_{fr'}-2\right),
\end{equation}
the Ising energy function becomes
\begin{equation}
    \frac{1}{2}\sum_{r=1}^{|R|}\sum_{r'=1}^{|R|}J_{rr'}s_rs_{r'}+\sum_{r=1}^{|R|}h_rs_r + \mathrm{const}.
\end{equation}
where the constant is equal to $\frac{1}{4}\sum_{f=1}^{|F|}\left(\sum_{r=1}^{|R|}a_{fr}-2\right)^2$. The sum of all the diagonal terms $(i=j)$ in the first sum is equal to $\mathrm{Tr}(J)$ since $s_i^2=1$; because $J_{ij}$ is symmetric i.e. $J_{ij}=J_{ji}$, we can further simplify the expression and write the Ising energy function as
\begin{equation}
    \label{eq:EnergyFunctionFinal}
    \mathcal{E}(s_1,\ldots,s_{|R|})=\sum_{r<r'}J_{rr'}s_rs_{r'}+\sum_{r=1}^{|R|}h_rs_r+\mathrm{const},
\end{equation}
where we have absorbed $\frac{1}{2}\mathrm{Tr}(J)$ into the constant. Finally, by promoting the spin variables to Pauli spin matrices $s_i\rightarrow\hat{\sigma}^z_i$, a cost Hamiltonian in the form of Eq.~\eqref{eq:CostFunction} is obtained.
%=============================================
% QUANTUM APPROXIMATE OPTIMIZATION ALGORITHM
%=============================================
\subsection{\label{app:QAOA}The Quantum Approximate Optimization Algorithm}
\begin{figure}[t]
    \includegraphics[width=\linewidth]{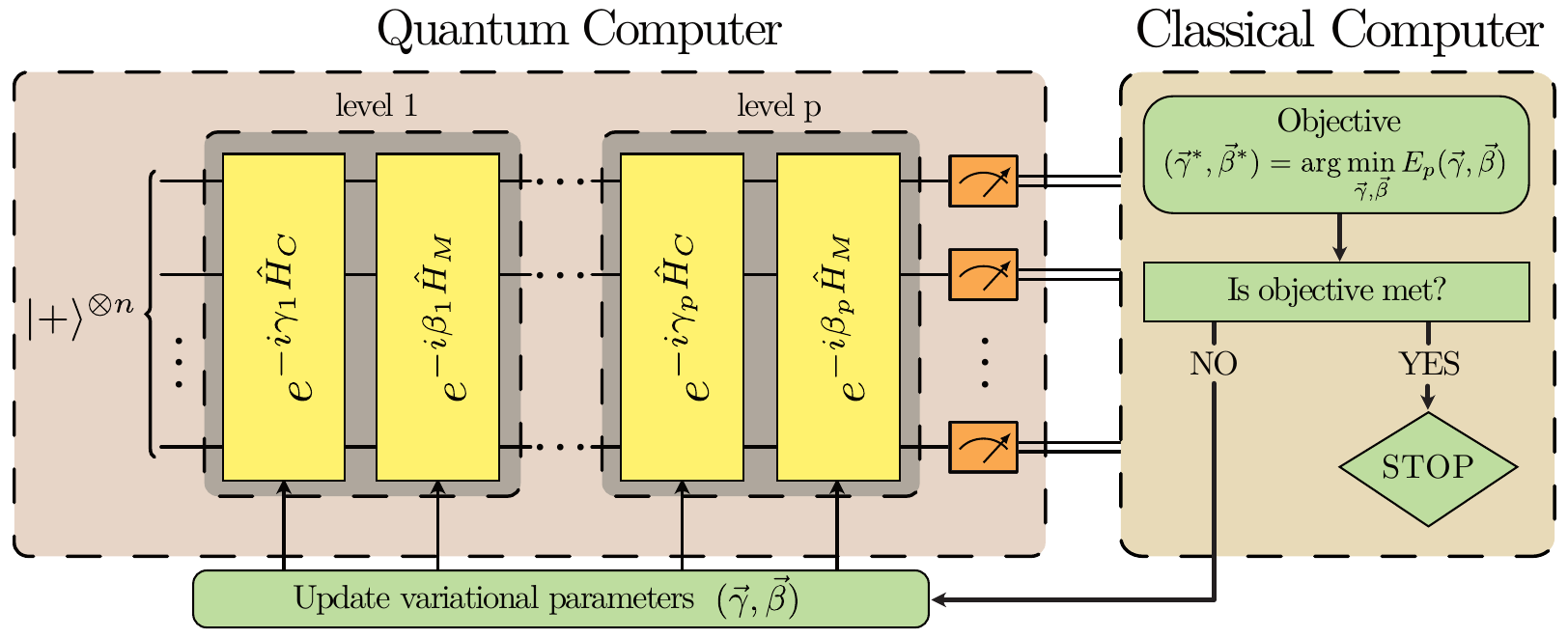}
    \caption{\label{fig:qaoaCircuit}Schematic representation of the QAOA. The quantum processor prepares the variational state, depending on variational parameters. The variational parameters $(\vec{\gamma},\vec{\beta})$ are optimized in a closed loop using a classical optimizer.}
\end{figure}

The QAOA starts from an initial quantum state which is taken as a superposition of all possible computational basis states $\ket{+}^{\otimes n}$. The second step of QAOA is to apply in an alternating sequence two parametrized non-commuting quantum gates, $\hat{U}(\gamma)$ and $\hat{V}(\beta)$, that are defined as:
\begin{align}
    \hat{U}(\gamma)\equiv e^{-i\gamma \hat{H}_C},\quad\hat{V}(\beta) \equiv e^{-i\beta \hat{H}_M},
\end{align}
where $\hat{H}_C$ is the cost Hamiltonian given by Eq.~\eqref{eq:CostFunction}, and $\hat{H}_M\equiv\sum_{i=1}^n\hat{\sigma}^x_i$ is a so called mixing Hamiltonian. The alternating sequence continues for a total of $p$ times with different variational parameters $\vec{\gamma}=(\gamma_1,\ldots,\gamma_p)$ with $\gamma_i\in[0,2\pi]$ if $\hat{H}_C$ has integer-valued eigenvalues, and $\vec{\beta}=(\beta_1,\ldots,\beta_p)$ with $\beta_i\in[0,\pi]$, such that the final variational state obtained is:
\begin{equation}
    \label{eq:FinalState}
    \ket*{\psi_p(\vec{\gamma},\vec{\beta})} \equiv \hat{V}(\beta_p)\hat{U}(\gamma_p)\ldots \hat{V}(\beta_1)\hat{U}(\gamma_1)\ket{+}^{\otimes n}.
\end{equation}
The parametrized quantum gates are then optimized in a closed loop using a classical optimizer, see Fig.~\ref{fig:qaoaCircuit}. The objective of the classical optimizer is to find the optimal variational parameters that minimize the expectation value of the cost Hamiltonian
\begin{equation}
    \label{eq:BestAngles}
    (\vec{\gamma}^*,\vec{\beta}^*)=\mathrm{arg}\min_{\vec{\gamma},\vec{\beta}}E_p(\vec{\gamma},\vec{\beta}),
\end{equation}
where
\begin{equation}
    \label{eq:ObjectiveFunction}
    E_p(\vec{\gamma},\vec{\beta})\equiv\matrixelement*{\psi_p(\vec{\gamma},\vec{\beta})}{\hat{H}_C}{\psi_p(\vec{\gamma},\vec{\beta})}.
\end{equation}
Note that this requires in principle multiple state preparations and measurements. Once the best possible variational parameters are found, they are used to create the state $\ket*{\psi_p(\vec{\gamma}^*,\vec{\beta}^*)}$, using the quantum processor for the state preparation. Then, one samples from this state by measuring in the computational basis, and the cost of the configuration obtained in the measurement, given by Eq.~\eqref{eq:CostFunction}, is evaluated. The latter step is classically efficient.

%================BEGIN FIGURE=================
%=================END FIGURE==================

The success probability is defined as the probability of finding the qubits in their ground state configuration $\ket*{x_\mathrm{sol}}$ when performing a single shot measurement of the $\ket*{\psi_p(\vec{\gamma},\vec{\beta})}$ state, i.e.
\begin{equation}
    \label{eq:Fidelity}
    F_p(\vec{\gamma},\vec{\beta}) \equiv |\braket*{x_\mathrm{sol}}{\psi_p(\vec{\gamma},\vec{\beta})}|^2,
\end{equation}
where $x_\mathrm{sol}=x_1x_2\ldots x_n$ is the bit string corresponding to the solution. Given this success probability we can ask: what is the probability of having observed the solution at least once after $m$ repeated measurements? The answer is given by:
\begin{equation}
    1-(1-F_p(\vec{\gamma},\vec{\beta}))^m.
\end{equation}
Thus to have the probability $(1-\varepsilon)$ of observing the solution, $m$ has to be
\begin{equation}
    \label{eq:BoundOnm}
    m > \frac{\log{\varepsilon}}{\log{(1-F_p(\vec{\gamma},\vec{\beta}))}}.
\end{equation}
To fix the ideas, consider a fair coin. In order to have a probability higher than $99.9~\%$ of observing Head at least once, one has to flip and ``measure" the coin $10$ times.

In what follows, we are going to apply this paradigm to solve the Exact Cover problem, by using the corresponding cost Hamiltonian, expressed by Eq.~\eqref{eq:CostFunction} with $J_{ij}$ and $h_i$ given by Eq.~\eqref{eq:Interaction} and \eqref{eq:MagneticField} respectively.
%=============================================
%                   RESULTS
%=============================================
\section{\label{sec:Results}Results}
We will examine instances for three different problem sizes of the Tail Assignment problem given in Table~\ref{tab:Instances}, corresponding to $8$, $15$ and $25$ routes. As clear from Eq.~\eqref{eq:CostFunction}, this requires quantum processors with $8$, $15$ and $25$ qubits respectively. 
%=================BEGIN TABLE=================
\begin{table}[h]
    \centering
    \caption{\label{tab:Instances}Information about the problem instances.}
    \begin{ruledtabular}
    \begin{tabular}{cccc}
        Routes & Flights & No. of instances & No. of sol. per instance \\ \hline
        8   & 77    & 10 & 1 \\
        15  & 77    & 9  & 1 \\
        25  & 278   & 10 & 1
    \end{tabular}
    \end{ruledtabular}
\end{table}
%==================END TABLE==================
%=============================================
%               ENERGY LANDSCAPE
%=============================================
\subsection{\label{sec:EnergyLandscape}Energy landscape}
Firstly, we can reduce the search space by noting that the eigenvalues of both Hamiltonians $\hat{H}_C$ and $\hat{H}_M$ are integer-valued. As a consequence, the expectation value Eq.~\eqref{eq:ObjectiveFunction} has even-symmetry, i.e. $E_p(\vec{\gamma},\vec{\beta})=E_p(-\vec{\gamma},-\vec{\beta})$. This symmetry allow us to restrict the domain of each $\gamma_i$ to $\gamma_i\in[0,\pi]$.

\begin{figure}[ht]
    \includegraphics[width=\linewidth]{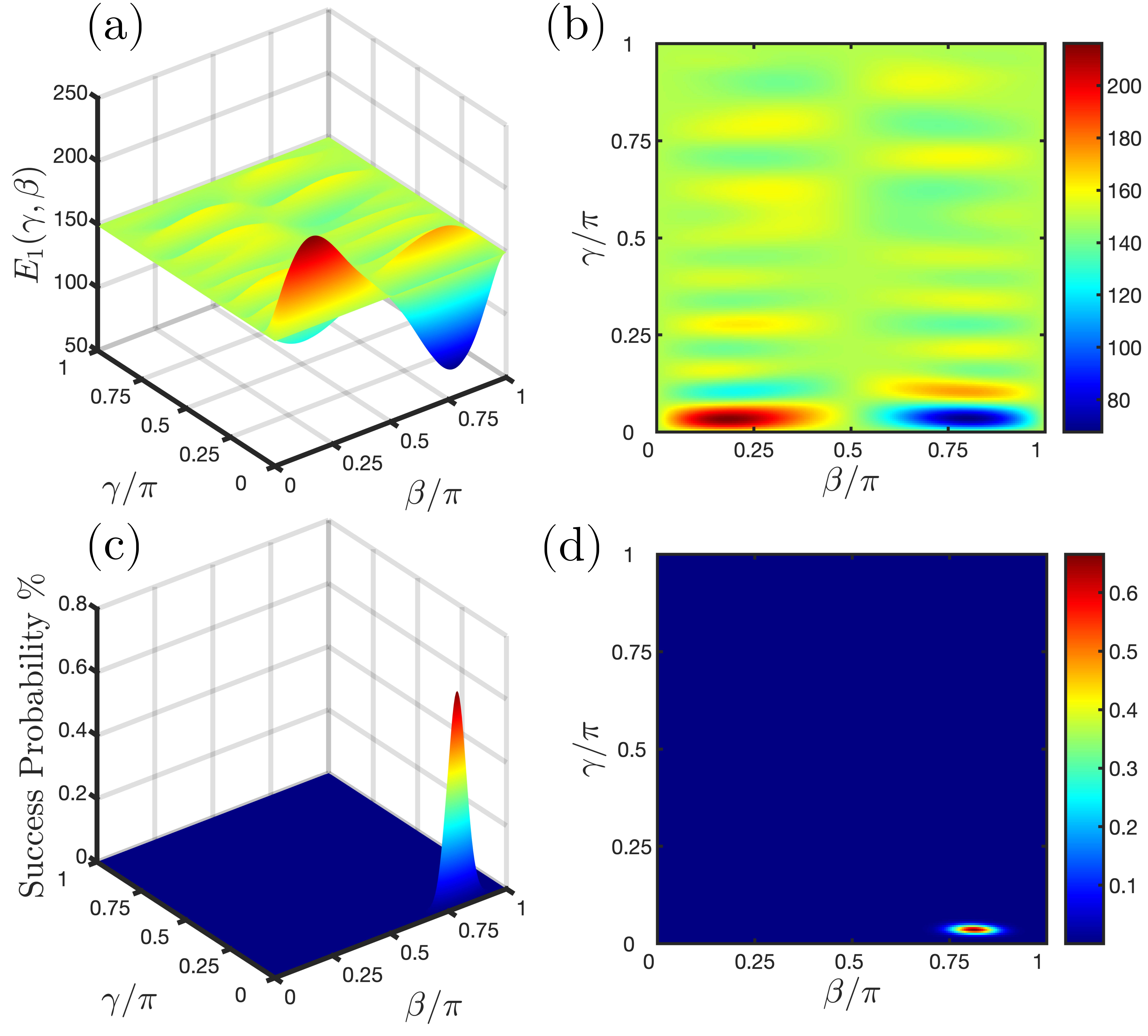}
    \caption{\label{fig:BruteForce25}(color online) Simulation results for one of the $25$ route instances as a function of $\gamma$ and $\beta$ for $p=1$. (a) and (b) Expectation value $E_1(\gamma,\beta)$; (c) and (d) Success probability $F_1(\gamma,\beta)$.}
\end{figure}

To highlight the difficulty of finding the best variational parameters we can visualize the landscape of the expectation value $E_1(\gamma,\beta)$, as well as the corresponding success probability $F_1(\gamma,\beta)$, as a function of $\gamma$ and $\beta$, for $p=1$, by evaluating them on a fine grid $\left[0,\pi\right]\times\left[0,\pi\right]$. Fig.~\ref{fig:BruteForce25} shows the simulation result for one of the $25$ route instances. The variational parameters resulting in the lowest expectation value, $(\gamma_\text{exp},\beta_\text{exp})$, and those resulting in the highest success probability, $(\gamma_\text{succ},\beta_\text{succ})$, are approximately the same. In fact $|\gamma_\text{exp}-\gamma_\text{succ}|\simeq 0$ and $|\beta_\text{exp}-\beta_\text{succ}|\simeq 0.047$. Note that this is not obvious, since QAOA only minimizes the expectation value, and does not explicitly maximize the success probability; a low expectation value does not necessarily translates onto a high success probability. For example, consider a variational state that is a linear combination of low energy excited eigenstates of the cost Hamiltonian. This state could potentially have a low expectation value while the success probability is zero. Similarly, a variational state that is a linear combination of the ground state with high energy eigenstates could have a high success probability, while the cost Hamiltonian expectation value is large. However, in the limit $p\rightarrow \infty$, $100~\%$ success probability is always achieved~\cite{Farhi2014}. For our problem, it is clear from Eq.~(\ref{eq:PenalizedFlightCover}) that the minimum energy of the first excited state is at least 1, so if we find an average cost which is lower than 1 for our variational state, we know that the ground state is a part of this state. The corresponding plots for one of the $8$ and $15$ route instances are shown in Appendix~\ref{app:Figures}. We note that all figures have qualitatively similar shape and that the optimal variational parameters for $p=1$ are located in the same region.
%================BEGIN FIGURE=================
%=================END FIGURE==================
%=============================================
%                   PATTERNS
%=============================================
\subsection{\label{sec:Patterns} Low iteration levels: Patterns in optimal variational parameters}
Before we look at the performance of QAOA, we will search for patterns in the optimal variational parameters for low iteration levels of the QAOA algorithm, namely up to $p=5$. Patterns in the optimal variational parameters have been observed before in the context of Max-Cut in Ref.~\cite{Zhou2018}, where it was shown that if a pattern exist it is possible to use different heuristics that can drastically speed up the classical optimization part of QAOA. This can potentially help us simulate the solution of our instances for intermediate $p$-level beyond $p = 5$, namely for $5 < p \leq 20$.

In order to find the optimal variational parameters, one possible approach would consist of a grid search method. However, evaluation of the cost Hamiltonian expectation value on a fine grid for higher dimensions quickly becomes computational expensive due to the large search space $\left[0,\pi\right]^p\times\left[0,\pi\right]^p$. Therefore, we discard the grid search method and resort to another optimization routine for finding good variational parameters for $1\leq p\leq 5$. This optimization routine is still exhaustive but more computational efficient. It distributes several random start points in the variational parameter landscape, and runs the gradient based BFGS algorithm~\cite{Broyden1970, *Fletcher1970, *Goldfarb1970, *Shanno1970} for every start point from which it records the global optimum. We provide relevant details in  Appendix~\ref{app:Simulations}. In Fig.~\ref{fig:BestAngles8} we present the optimal variational parameters $(\vec{\gamma}^*,\vec{\beta}^*)$ from $p=3$ up to $p=5$ for the $8$ route instances. We observe that a persistent pattern shows up, and that both $\gamma_i$ and $\beta_i$ tend to increase slowly with $i=1,2,\ldots,p$. An analogous analysis for the $15$ route instances, shown in the Appendix in Fig.~\ref{fig:BestAngles15}, yields a qualitatively similar result. For the $25$ route instances, it was not possible to perform this analysis, because for $p>1$ performing an exhaustive search becomes too computationally expensive.
%================BEGIN FIGURE=================
\begin{figure}[H]
    \includegraphics[width=\linewidth]{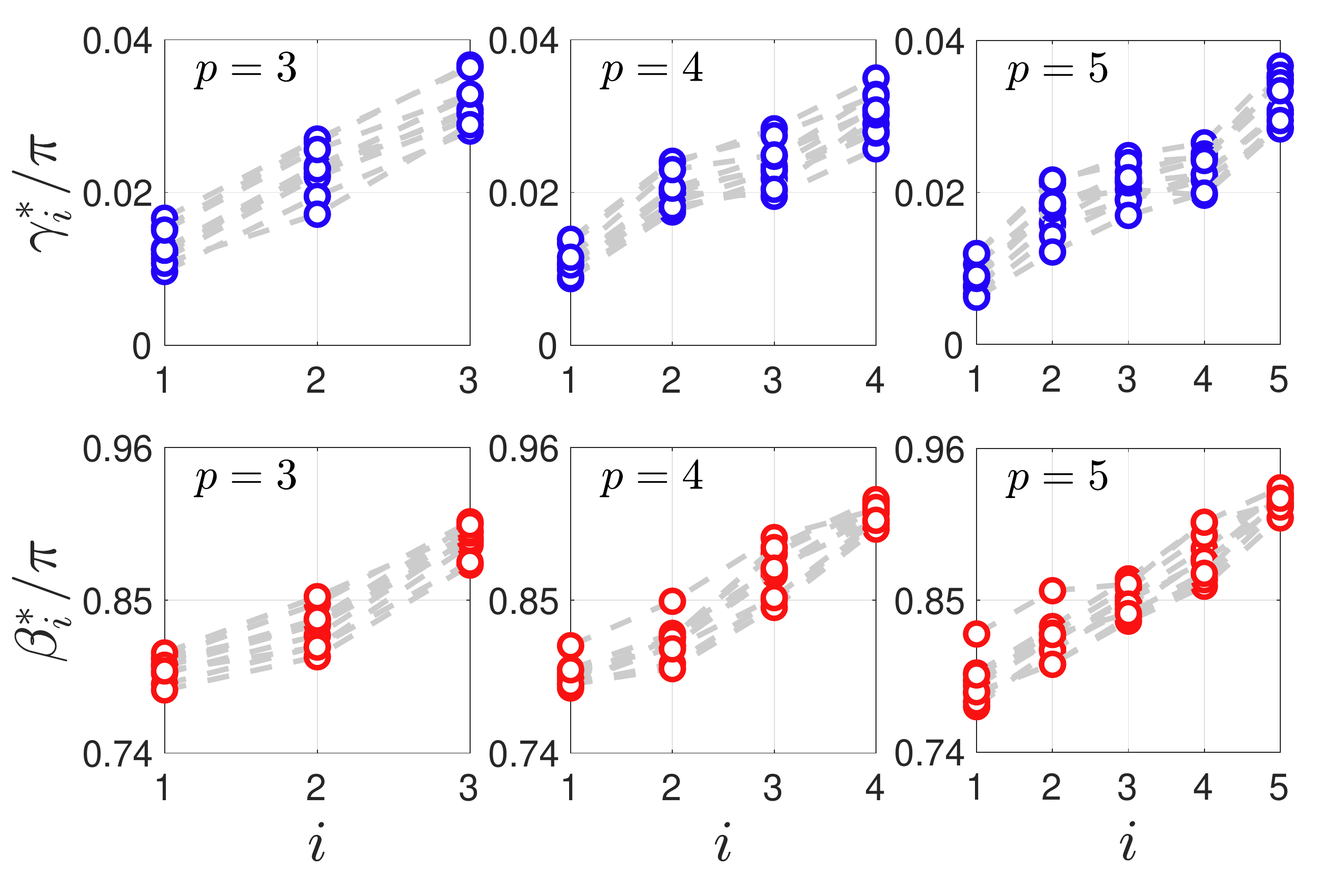}
    \caption{\label{fig:BestAngles8}
    The optimal QAOA variational parameters $(\vec{\gamma}^*,\vec{\beta}^*)$ for the $8$ route instances, for $3\leq p \leq 5$. The pattern is visualized by plotting the optimal variational parameters where each gray dashed line connects the variational parameters for one $8$ route instance.}
\end{figure}
%=================END FIGURE==================

%================BEGIN FIGURE=================
\begin{figure*}[t!]
    \includegraphics[width=1\linewidth]{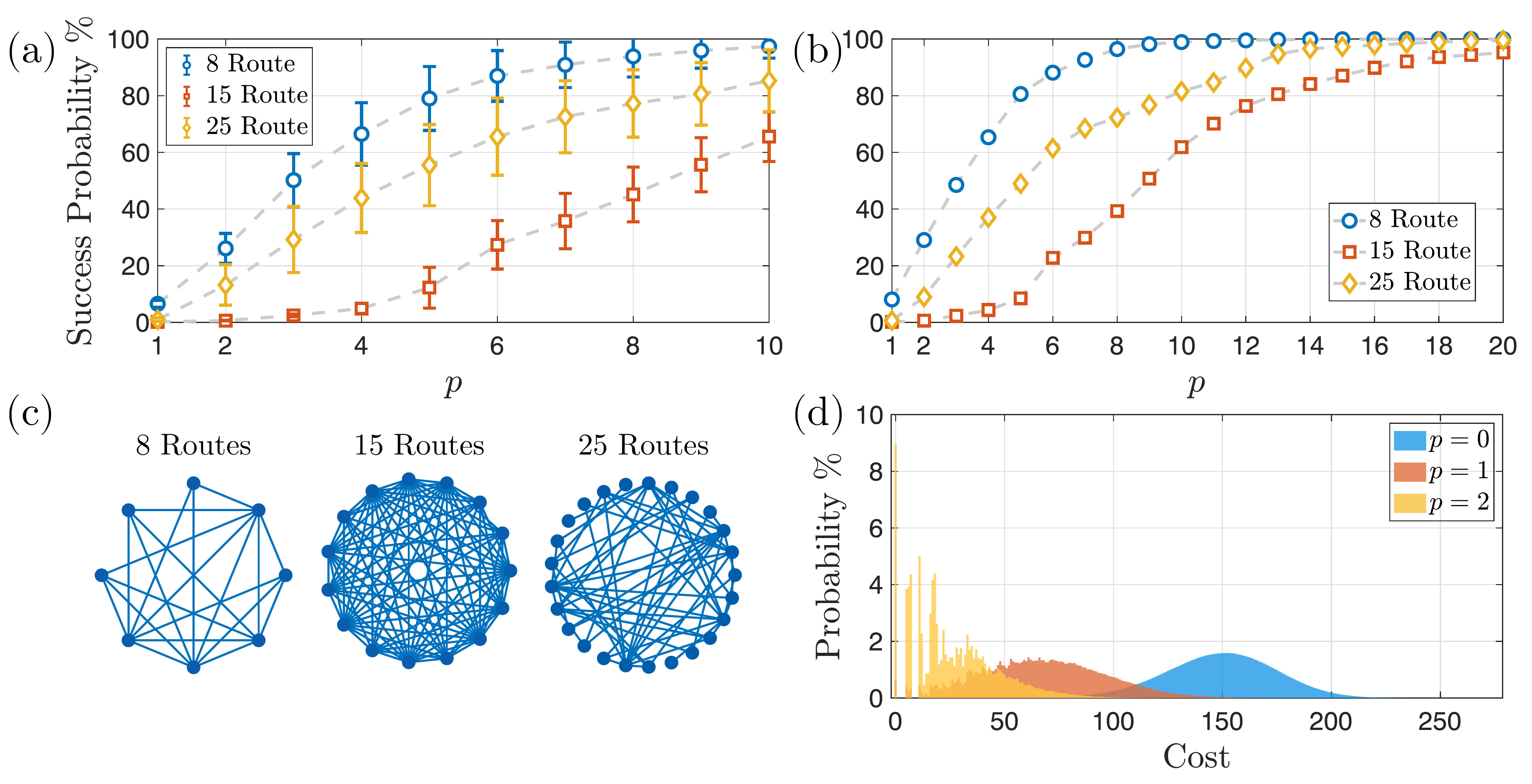}
    \caption{\label{fig:Results}(color online) (a) Average success probability as a function of the iteration level $p$ using the best found variational parameters for the three different problem sizes. The error-bars in the figure represent the standard deviation of the average success probability. (b) Success probability $F_p(\vec{\gamma}^*,\vec{\beta}^*)$ as a function of $p$ for one selected instance from each problem size. (c) Graph representation of the three instances shown in (b). (d) Probability that a measurement of the state $\ket*{\psi_p(\vec{\gamma}^*,\vec{\beta}^*)}$ will yield a certain cost (or equivalently, eigenvalue of the cost Hamiltonian) for the iteration levels $p=0,1,2$, where $p=0$ is the initial or ``random'' state $\ket*{+}^{\otimes n}$.}
\end{figure*}
%=================END FIGURE==================
%=============================================
%               SUCCESS PROBABILITY
%=============================================
\subsection{\label{sec:SucccessProbability} Intermediate iteration levels: Analysis of success probability}
Based on the patterns found in the previous section, we now use an interpolation-based strategy, introduced in~\cite{Zhou2018}, in order to study the performance of intermediate $p$-level QAOA. This strategy consists in predicting a good starting point for the variational parameters search at level $p+1$ for each individual instance based on the best variational parameters found at level $p$ for the same instance. From the produced starting-point we run the gradient-free Nelder-Mead method~\cite{Nelder1965,Lagarias98}, which is reported in Ref.~\cite{Zhou2018} to work equally well as the BFGS method, for this heuristic strategy. The Nelder-Mead algorithm was implemented in MATLAB version R2019b using the \texttt{fminsearch} function. Furthermore, in order to force the Nelder-Mead algorithm to terminate after sufficiently many iterations, we set the two stopping criteria - maximum number of function evaluations and iterations - both to $60p$. We furthermore make the assumption that a pattern in the variational parameters also exists in each of the $25$ route instances, and we therefore use the interpolation strategy mentioned above for each of these instances as well, as an educated guess. We base this assumption on the qualitatively similar shape of the expectation value landscape that the three different problems sizes investigated had for $p=1$.

We use the aforementioned interpolation-strategy for finding good local optimal variational parameters up to $p=10$ for all the instances. The success probability as a function of iteration level $p$ averaged over all the instances for the three different problem sizes is plotted in Fig.~\ref{fig:Results}(a). Moreover, we select one instance from each problem size, for which we perform simulations up to $p=20$. In Fig.~\ref{fig:Results}(b) we plot the success probability for these three instances. The corresponding variational parameters $\vec{\gamma}^*$ and $\vec{\beta}^*$ are provided in Appendix~\ref{app:Figures}, Fig.~\ref{fig:VariationalParameters}. It is observed that the success probability increases with the parameter $p$ in both the averaged and the single-instance cases, reaching almost $100~\%$ for the instances where we have used high iteration level $p=20$. 

From the results in Figs.~\ref{fig:Results}(a) and \ref{fig:Results}(b) we also note that the $25$ route instances are easier to solve than the $15$ route instances, in the sense that the success probability is higher for the former instances at any given iteration level $p$ of the algorithm. This fact can seem counter-intuitive, as one could naively think that larger instances correspond to harder problems. We perform further analysis in order to explain this apparent contradiction. 

We start by representing each instance as a graph, by identifying $J_{ij}$ in Eq.~\eqref{eq:Interaction} with an adjacency matrix. In this way, each vertex in the graph represents a route and two vertices are connected by an edge if they share a flight. The valency of a vertex, i.e. the number of incident edges to the vertex, indicates how many ``clauses'' the vertex is contained in, or in other words how many other vertices it has to compete with. In Table~\ref{tab:Valency} we list the average valency of each vertex for the three problem sizes. We note that the $15$ instances have more than twice the average valency compared to the $25$ route instances. This is also visualized in Fig.~\ref{fig:Results}(c), where the graph connectivity for one instance of each problem size is represented. It is clear that the connectivity for the $15$ instance is the most dense. Establishing a general connection between the hardness of the instances and their valency is beyond the scope of our paper. However, such a connection is known to exists in some specific contexts, e.g. for Exact Cover by 3-sets~\cite{kalapala2005phase,raymond2007phase}. This hints to the fact that such a connection might exist also for our instances, despite they are not in the form of Exact Cover by 3-sets.
%=================BEGIN TABLE=================
\begin{table}[t] 
    \centering
    \caption{\label{tab:Valency}Valency of the graphs. The first column in the table is the number of routes. The second column is the average valency of a vertex taken as an average over all the instances. The corresponding standard deviation is given in the third column.}
    \begin{ruledtabular}
    \begin{tabular}{ccc}
        Routes & Mean & Standard deviation \\ \hline
        8   & 5.15   & 0.24 \\
        15  & 12.62  & 0.42 \\
        25  & 5.54   & 0.77
    \end{tabular}
    \end{ruledtabular}
\end{table}
%==================END TABLE==================
To elucidate further why denser graphs are more difficult to solve with the QAOA we recall, following Refs.~\cite{Farhi2014,Brandao2018}, that the expectation value Eq.~\eqref{eq:ObjectiveFunction} can be expressed as a sum of expectation values involving all possible subgraphs. Subgraphs are obtained by starting from an edge $\langle ij\rangle$ of a graph, e.g. the type of graph given in  Fig.~\ref{fig:Results}(c), and ``walking" along the graph at most $p$ steps away from that edge, for a given iteration level $p$. Indicating with $f_g(\vec{\gamma}, \vec{\beta})$ the contribution to the expectation value from subgraph $g$, and with $w_g$ the corresponding subgraph occurrence, it is possible to re-write the expectation value as $E_p(\vec{\gamma}, \vec{\beta}) = \sum_g w_g f_g(\vec{\gamma}, \vec{\beta})$. Since the contribution to the expectation value is different for each subgraph, the higher the number of important subgraphs (with a significant $w_g$) is, the harder it will be to make the cost close to zero for a given iteration level $p$, since the QAOA need to make each individual term in the sum small. Since the average valency of a graph contributes to the number of subgraphs, this results in a lower success probability for the $15$ route instances, as (as we have shown in Fig.~\ref{fig:Results}(c) and Table~\ref{tab:Valency}) those possess higher average valency.

Finally, in Fig.~\ref{fig:Results}(d) we visualize how the probability of measuring a certain cost, or equivalently an eigenvalue of the cost Hamiltonian, given the state $\ket*{\psi_p(\vec{\gamma}^*,\vec{\beta}^*)}$, changes for each iteration $p=0,1,2$ of QAOA using the best found variational parameters for one of the $25$ route instances. It is clear that the effect of iterating QAOA is that the probability of configurations with lower cost increases. This validates the effectiveness of QAOA in producing output configurations corresponding to low energy states of the cost Hamiltonian, when the iteration level $p$ is increased. In particular, for $p=2$ a peak at the zero-cost configuration appears clearly, corresponding to a success probability of $8.97~\%$. This results in only $74$ measurements needed, in order to have a probability greater than $99.9~\%$ of measuring the solution at least once.

In order to benchmark the effectiveness of QAOA in solving this problem against other quantum algorithms, in Appendix~\ref{app:Comparision} we compare the time to solution of QAOA with that of quantum annealing, and find that QAOA outperforms quantum annealing for all the $8$ and $15$ route instances.

Finally, noise and imperfection in practical experimental implementations on a quantum computer will induce departures from the obtained success probabilities, and it is an open question whether realistic hardware will still be able to produce the good solution, with satisfactory success probability. Although a complete study of the effect of noise is beyond the scope of the present paper, in Appendix~\ref{app:Noise} we characterize the effect of a simple depolarizing noise model, to study how noise affects the performance of QAOA. As expected, we find that with noise an optimal value of $p$ exists. Beyond that value of $p$, the success probability starts to decrease, due to the larger effect of decoherence when the gate sequence becomes longer. However, for the optimal $p$, the success probability is only halved, still pointing to relevance of the use of QAOA for solving this problem even in realistic experimental conditions.
%=============================================
%                 CONCLUSIONS
%=============================================
\section{\label{sec:Conclusions}Conclusions}
In conclusions, we have simulated the solution of instances of the Exact Cover problem that stem as a reduction of the Tail Assignment problem to the case where the goal is to find any solution satisfying all the constraints, using the QAOA.

Our results indicate that these instances can be solved satisfactorily by means of QAOA, yielding relative high success probabilities even for low iteration level of the algorithm. For instance, for the $25$ qubits case we obtain a success probability of $8.97~\%$ for $p=2$ in the single measurement scenario. This corresponds to a success probability of $99.9~\%$ for $74$ repeated measurements. This low iteration level translates into a low circuit depth needed for the implementation of this algorithm, corroborating feasibility on a near-term quantum device.

Moreover, we observed patterns for the variational parameters $(\vec{\gamma},\vec{\beta})$ which allowed for a substantial simplification of the classical optimization problem of finding the best variational parameters, despite the fact that the problem instances have been extracted from a real world problem.

Our analysis has revealed non-trivial properties in the connectivity of the instances considered. I.e., the $15$ qubit instances were more connected than the $25$ qubit ones. A thorough study of the connectivity and graph-type that are relevant for the Tail Assignment problem in the context of complex quantum networks~\cite{Sansavini2019,Albert2002} is beyond the scope of the present paper, but stems as an interesting perspective. Another interesting question is whether the implementation of the QAOA algorithm on hardware with restricted connectivity would still yield non-trivial success probabilities, as shown in Ref.~\cite{farhi2017quantum} for Max-Cut on three-regular graphs.

Our successful solution with QAOA of small-size instances of Exact Cover extracted from Tail Assignment  motivates further studies, such as the use of QAOA for solving instances with multiple feasible solutions, where costs are re-introduced, and where the number of considered routes is larger, towards tackling real-world instances.

It remains  an open question how the performance of QAOA compares with existing classical algorithms for solving large instances of the Exact Cover problem extracted from the Tail Assignment problem.However, we expect that current known methods as Branch-and-Bound,  Cutting planes or Branch-and-Cut \cite{Conforti_2014} will perform well on these small instances. Further investigations are needed in order to compare the scaling in terms of time complexity of QAOA fixing a target success probability (i.e. the required iteration level $p$) and standard classical methods, when the size of the problem increases.

While finalizing this work, we became aware of an alternative method for the optimization of the variational parameters, that makes use of the \textit{Gibbs objective function}, defined as $-\log \expectationvalue*{e^{-\eta \hat{H}_C}}$, where $\eta>0$, instead of the expectation value Eq.~\eqref{eq:CostFunction}~\cite{li2019}. This approach is expected to be superior because the Gibbs objective function rewards lower energy states, which increases the success probability. We leave the use of this approach for optimization of the variational parameters in our problem to further study.

%
% ****** Acknowledgments ******
\acknowledgments
We thank Devdatt Dubhashi and Eleanor Rieffel for fruitful discussions. This work was supported from the Knut and Alice Wallenberg Foundation through the Wallenberg Center for Quantum Technology (WACQT). G. F. acknowledges financial support from the Swedish Research Council through the VR project QUACVA.

%
% ****** References ******
\bibliographystyle{apsrev4-1}
\bibliography{references}

%
% ****** Appendix ******
\appendix

%=============================================
%               ADDITIONAL FIGURES
%=============================================
\section{\label{app:Figures}Additional figures}
%================BEGIN FIGURE=================
\begin{figure}[H]
    \includegraphics[width=\linewidth]{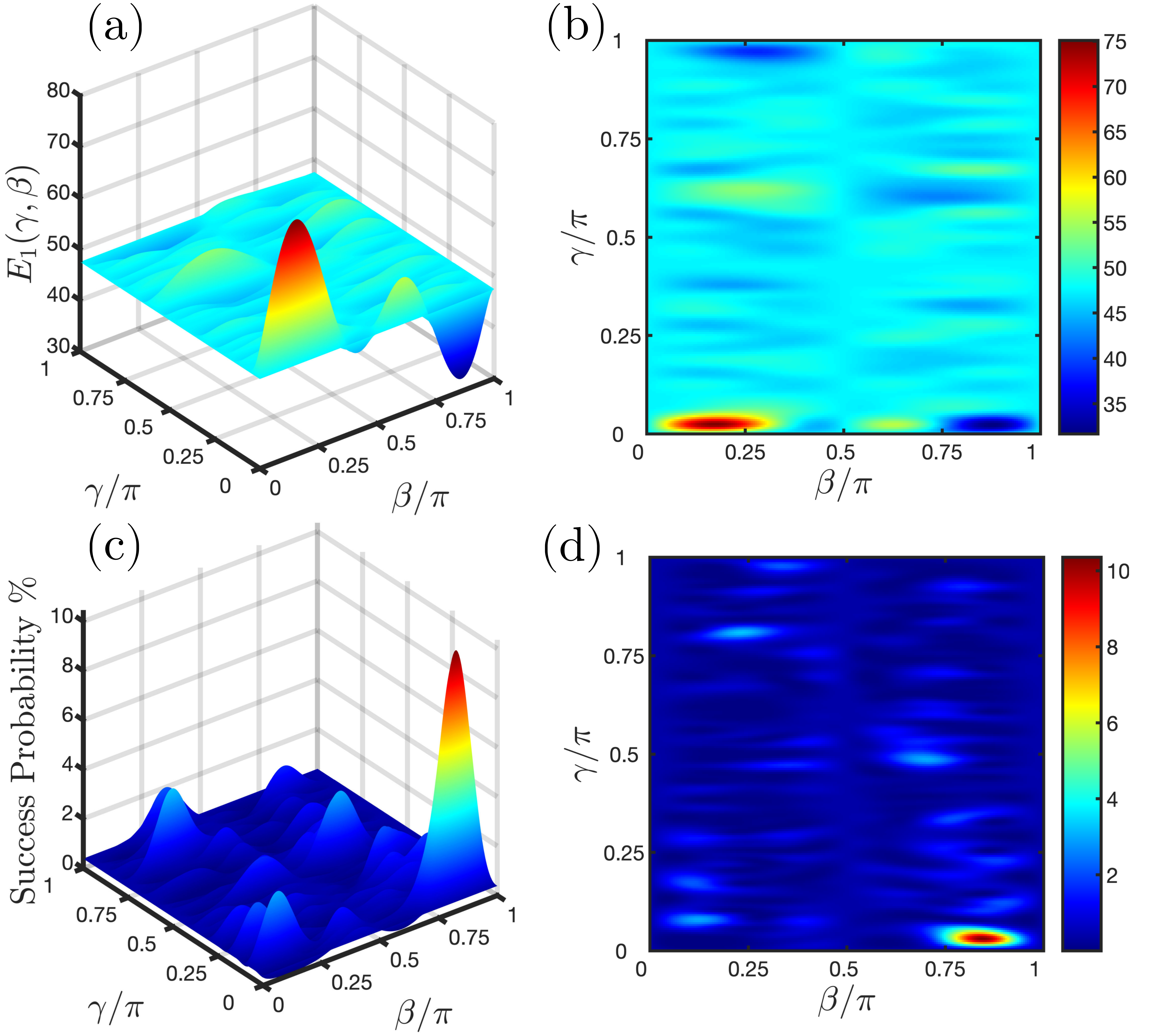}
    \caption{\label{fig:BruteForce8} (color online) Simulation results for one of the $8$ route instances as a function of $\gamma$ and $\beta$ for $p=1$. (a) and (b) Expectation value $E_1(\gamma,\beta)$; (c) and (d) Success probability $F_1(\gamma,\beta)$.}
\end{figure}
%=================END FIGURE==================

%================BEGIN FIGURE=================
\begin{figure}[H]
    \includegraphics[width=\linewidth]{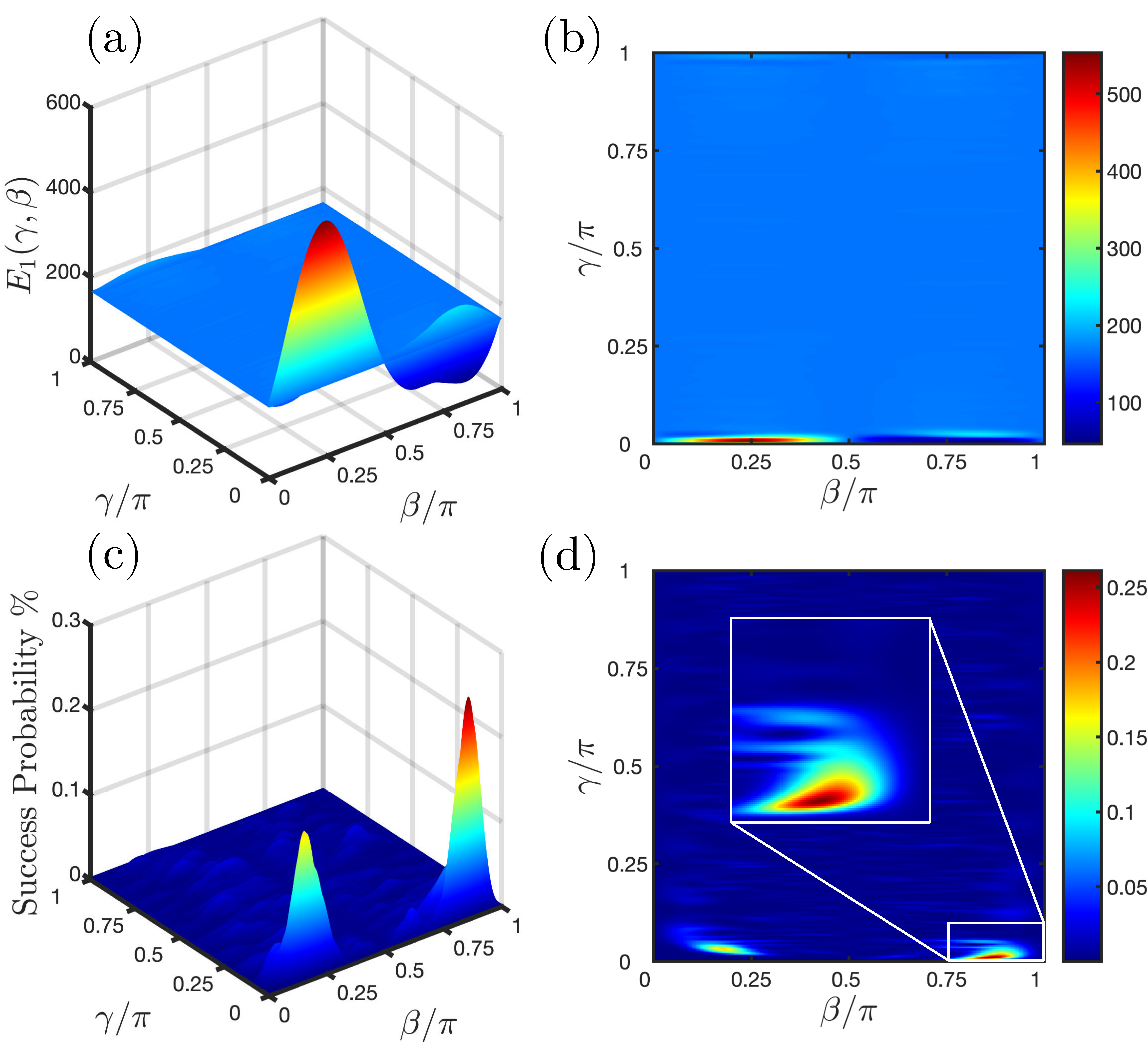}
    \caption{\label{fig:BruteForce15} (color online) Simulation results for one of the $15$ route instances as a function of $\gamma$ and $\beta$ for $p=1$. (a) and (b) Expectation value $E_1(\gamma,\beta)$; (c) and (d) Success probability $F_1(\gamma,\beta)$.}
\end{figure}
%=================END FIGURE==================

%================BEGIN FIGURE=================
\begin{figure}[H]
    \includegraphics[width=\linewidth]{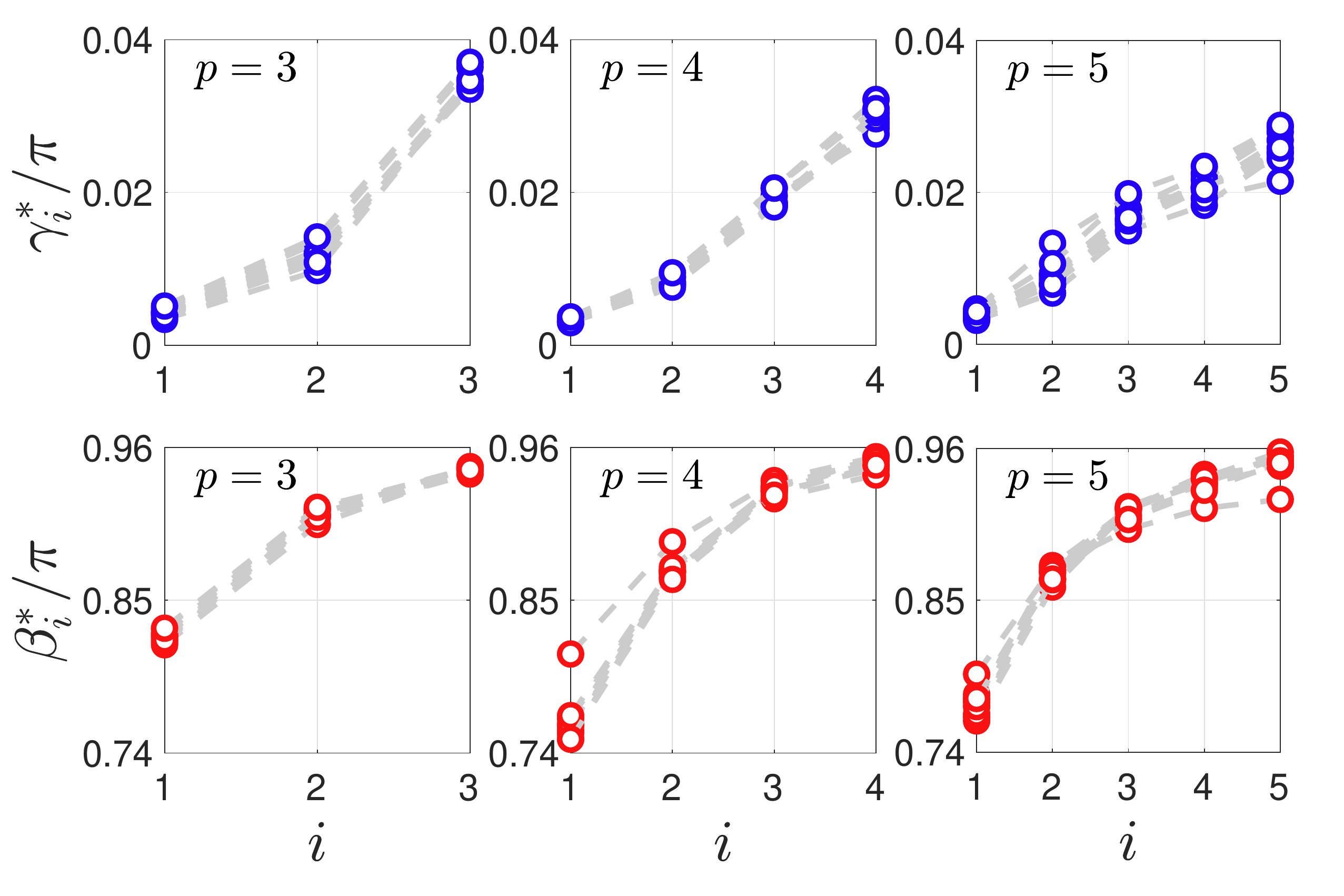}
    \caption{\label{fig:BestAngles15}The optimal QAOA variational parameters $(\vec{\gamma}^*,\vec{\beta}^*)$ for the $15$ route instances, for $3\leq p \leq 5$. The pattern is visualized by plotting the optimal parameters where each grey dashed line connects the optimal variational parameters of one particular instance.}
\end{figure}
%=================END FIGURE==================

%================BEGIN FIGURE=================
\begin{figure}[H]
    \includegraphics[width=\linewidth]{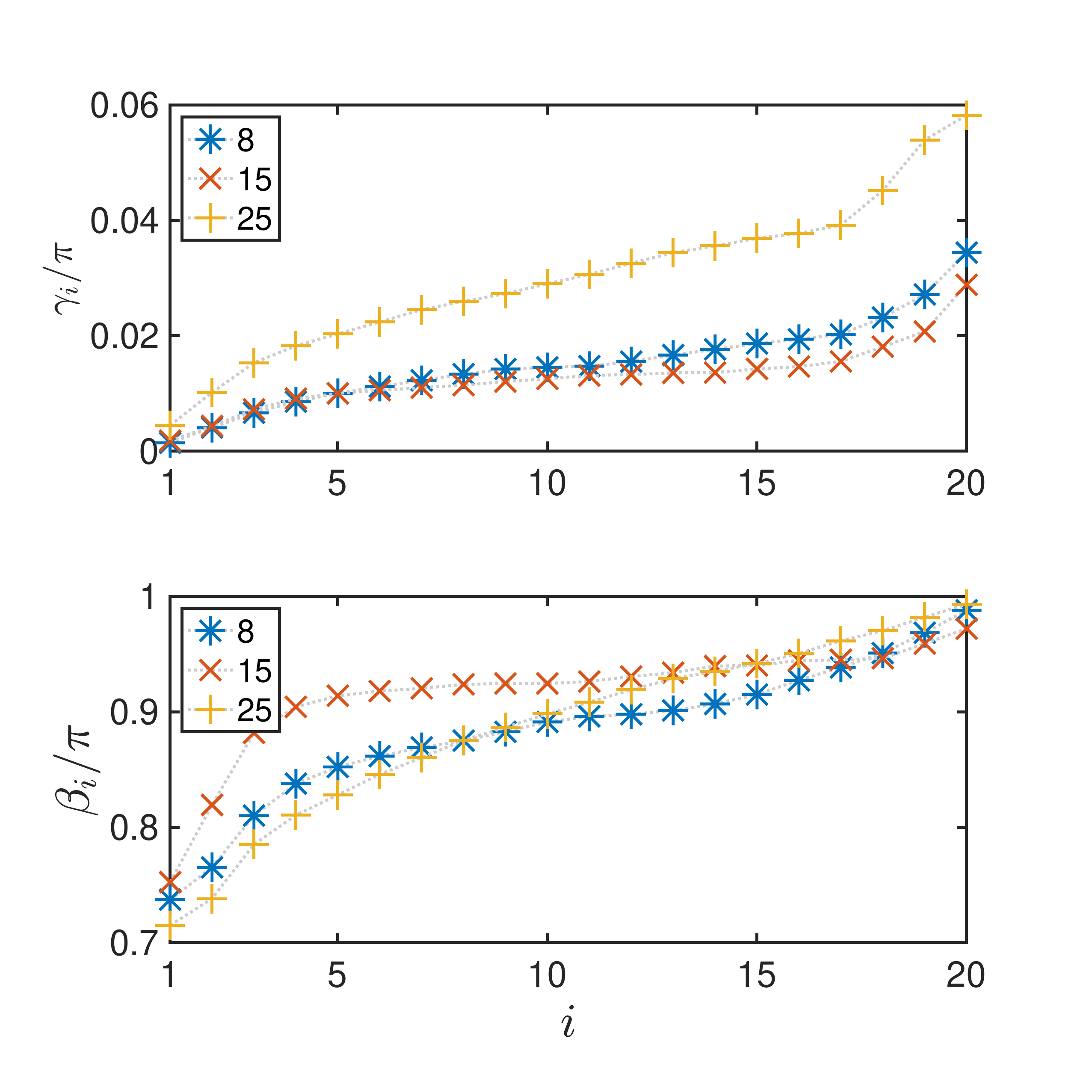}
    \caption{\label{fig:VariationalParameters}The best found $\vec{\gamma}^*$ and $\vec{\beta}^*$ for the three instances shown in Fig.~\ref{fig:Results}~(b).}
\end{figure}
%=================END FIGURE==================

%=============================================
%            NUMERICAL SIMULATIONS
%=============================================
\section{\label{app:Simulations}Numerical simulations}
The numerical simulations for the exhaustive search method was done in MATLAB version R2019b where the \texttt{MultiStart} function was used to search thoroughly for the optimal variational parameters. \texttt{MultiStart} attempts to find multiple local minimums to the objective function by starting from various points in the variational parameter landscape. When run, it distributes start points to multiple processors (cpus) that run in parallel. From a start point it runs a \emph{local solver} and when the solver reaches a stopping criterion it terminates and the obtained minima from the solver is stored in an array. When \texttt{MultiStart} runs out of start points it stops, and the array with minimums from the solver is sorted by the objective function value in ascending order. The parameters where the objective function is the lowest is then returned as output. As \emph{local solver} we used the BFGS algorithm~\cite{Broyden1970, *Fletcher1970, *Goldfarb1970, *Shanno1970} which is implemented as {\tt fmincon} in MATLAB. The number of random start points was chosen to be $4\times10^3$. This number was empirically determined by running the simulations a few times for this value and observing that the minimum of the objective function always converged to the same value and gave the same parameters. As mentioned the solver stops when the solver's stopping criteria is met. Two examples of such criterion's are the function tolerance and the step tolerance. The first one, the function tolerance, is a lower bound on the change in the value of the objective function during a step, that is if $|F_p(\vec{\gamma},\vec{\beta}) - F_p(\vec{\gamma}',\vec{\beta}')| < \text{\tt{FunctionTolerance}}$, the iteration ends. The second one, the step tolerance, is such that if the solver attempts to take a step that is smaller than $|\vec{\gamma}-\vec{\gamma}^\prime|^2+|\vec{\beta}-\vec{\beta}^\prime|^2 < \tt{StepTolerance}$, the iteration ends. Both {\tt StepTolerance} and {\tt FunctionTolerance} were set to their default values which was $10^{-6}$.
%=============================================
%           QUANTUM ANNEALING VS QAOA
%=============================================
\section{\label{app:Comparision}Comparison: Time to solution of Quantum Annealing versus QAOA}
In this section we compare the time to solution of the quantum annealing (QA) algorithm with that of the QAOA. In quantum annealing we start from the same initial state as the QAOA, which is in fact the ground state of the mixing Hamiltonian that we use in QAOA, but with a minus sign in front, $\hat H_M^\textrm{QA}\equiv -\hat H_M=-\sum \hat \sigma^x_i$. By adiabatically changing from the mixing Hamiltonian to the cost Hamiltonian the system will remain in its instantaneous ground state throughout the evolution, and end up in the ground state of the cost Hamiltonian. For a linear time dependence, the quantum annealing Hamiltonian is given by
\begin{equation}
    \label{eq:QAHamiltonian}
    \hat H(t) = \frac{t}{T}\hat H_C + \left(1-\frac{t}{T}\right) \hat H_M^\textrm{QA}, \quad 0\leq t \leq T,
\end{equation}
where $\hat H_C$ is the cost Hamiltonian, $\hat H_M^\textrm{QA}$ is the quantum annealing starting Hamiltonian, and $T$ is the total annealing time. It is known that rather than running the algorithm adiabatically, it can be advantageous to run the algorithm for a shorter time (non fully adiabatically). On the one hand, this yields to a finite  probability to excite higher energy states and decreases the success probability on a single run; on the other hand, since the annealing time $T$ is shorter, one can then increase the number of repetitions, yielding an increase of the total success probability, on several runs. Therefore, one can define the \emph{time to solution}, which is a measure of how quickly the algorithm can find the optimal solution. The time to solution for QA is defined by \cite{Albash2018}
\begin{equation*}
    \mathrm{TTS}_\mathrm{QA}(T) = T\frac{\log(1-p_\mathrm{d})}{\log(1-F_\mathrm{gs}(T))},
\end{equation*}
%================BEGIN FIGURE=================
\begin{figure}[t]
    \includegraphics[width=\linewidth]{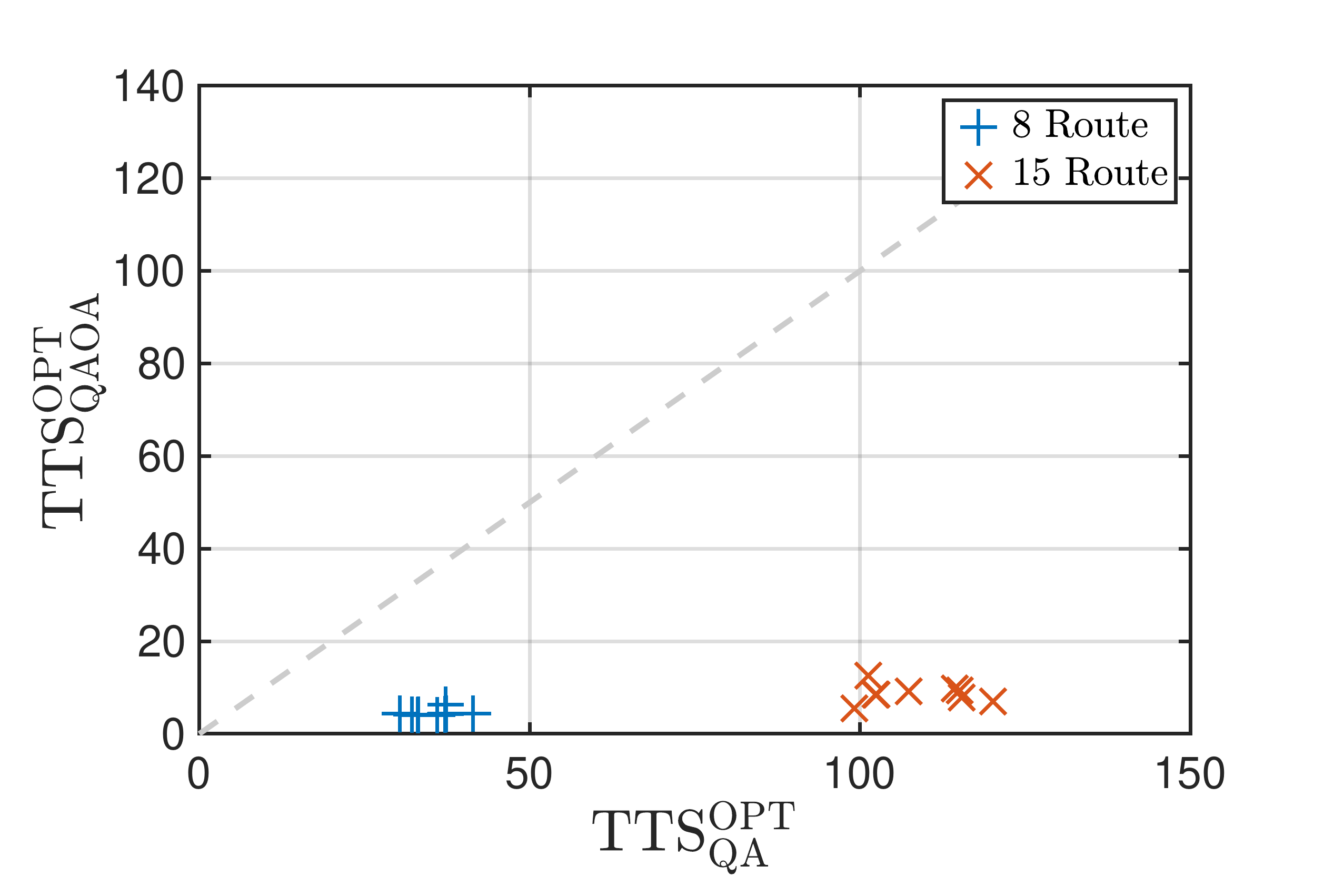}
    \caption{\label{fig:TTS}The optimal time to solution for QAOA and QA. The fact that the markers are below the dotted line means that QAOA outperforms QA in the time required to achieve a $99 \%$ success probability.}
\end{figure}
%=================END FIGURE==================
where $p_\mathrm{d}$ is the target success probability that we fix to $99~\%$, and $F_\mathrm{gs}(T)$ is the single shot success probability after running the algorithm for a time $T$. The optimal $\mathrm{TTS}_\mathrm{QA}(T)$ is thus given by the time $T$ that minimize
\begin{equation*}
    \mathrm{TTS}_\mathrm{QA}^\mathrm{OPT} = \min_{T>0} \mathrm{TTS}_\mathrm{QA}(T).
\end{equation*}
Following the spirit of Ref.~\cite{Zhou2018}, it is possible interpret the sum of the optimal variational parameters of the QAOA as the total ``annealing'' time that is used, in order to sequentially evolve the system under the action of each of the two Hamiltonians, $T_p = \sum_{i=1}^p(|\gamma^*_i|+|\beta^*_i|)$. Thus, the time to solution for QAOA is
\begin{equation*}
    \mathrm{TTS}_\mathrm{QAOA}(p) = T_p\frac{\log(1-p_\mathrm{d})}{\log(1-F_p(\vec{\gamma}^*,\vec{\beta^*}))},
\end{equation*}
where $F_p(\vec{\gamma}^*,\vec{\beta^*})$ is given by Eq.~\eqref{eq:Fidelity}.
Analogously as for QA, the optimal $\mathrm{TTS}_\mathrm{QAOA}(p)$ is given by
\begin{equation*}
    \mathrm{TTS}_\mathrm{QAOA}^\mathrm{OPT} = \min_{p>0}\mathrm{TTS}_\mathrm{QAOA}(p).
\end{equation*}
We would of course like $T_p$ to be as small as possible, therefore we subtract all the optimal $\beta^*$ values by $\pi$. We can do this since $\psi_p(\vec{\gamma},\vec{\beta})$ is $\pi$-periodic in $\beta$ up to a global phase. This $\pi$-shifted value of $\beta$ is the value that one would obtain, if one would choose to use the quantum annealing mixer Hamiltonian (i.e. the one with a minus in front of the summation), instead of the  mixer commonly used for the QAOA. 

We run the QA algorithm for all the $8$ and $15$ route instances for different total annealing times $T$ and record the optimal TTS that we find. In Fig.~\ref{fig:TTS} we plot the $\mathrm{TTS}^\mathrm{OPT}$ for both algorithms, and find that the $\mathrm{TTS}_\mathrm{QAOA}^\mathrm{OPT}$ is smaller than $\mathrm{TTS}_\mathrm{QA}^\mathrm{OPT}$ for all the instances. For   the $15$ route instances, QAOA is one order of magnitude faster in achieving 99 \% success probability.  

%=============================================
%                   NOISE
%=============================================
\section{\label{app:Noise}Depolarizing noise}
\begin{figure}[H]
    \includegraphics[width=\linewidth]{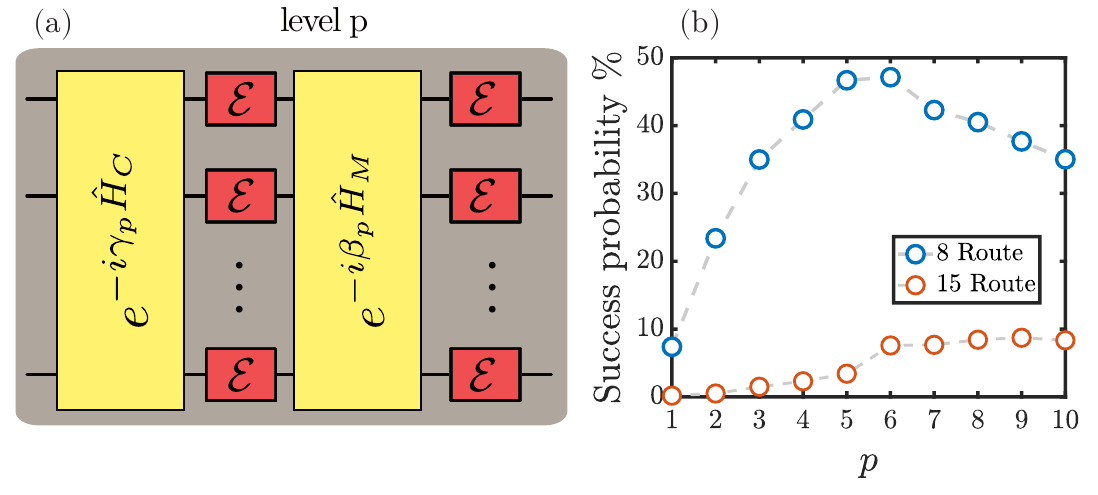}
    \caption{\label{fig:Noise}(a) After each application of the cost and mixing Hamiltonian of the QAOA an error gate $\mathcal{E}$ given by Eq.~\eqref{eq:Error} is independently applied to every qubit. (b) Success probability with noise for one of the $8$ and $15$ route instances.}
\end{figure}

In this Appendix we perform a simple study of how depolarizing noise affects the performance of QAOA. We model the depolarizing noise as random uncorrelated Pauli-$X$, $Y$ or $Z$ operations using the error gate
\begin{equation}
    \label{eq:Error}
    \mathcal{E} = (1-\eta)I + \frac{\eta}{3}(X+Y+Z), 
\end{equation}
where $\eta$ is the probability that an error occurs, that we fix to $1~\%$. This error gate acts on each individual qubit between the applications of the cost and mixing Hamiltonian, see Fig.~\ref{fig:Noise}(a). We then repeat the circuit sufficiently many times to get a statistical average over the noise. In Fig.~\ref{fig:Noise}(b) we plot the success probability with noise for the same $8$ and $15$ route instances as shown in Fig.~\ref{fig:Results}(b). A trade-off appears between the level of iteration of the algorithm $p$, and the success probability. In particular, we observe that for $p > 6$ the success probability starts to decrease for the $8$ route instance, while for the $15$ route instance it levels off, indicating that the gain of increasing one level $p$ equals the decrease due to the noise. This is expected, as faulty gates decrease the fidelity of the prepared state with the best theoretically found variational state. However, the resulting success probabilities at $p=6$ are roughly halved with respect to the noiseless case.

%================BEGIN FIGURE=================
%=================END FIGURE==================
\end{document}